# Deep Learning-Based Intrusion Detection System for Advanced Metering Infrastructure

Zakaria El Mrabet
STRS Lab
National Institute of Posts and Telecommunication
Rabat, Morocco
zakaria.elmrabet2@gmail.com

Mehdi Ezzari
STRS Lab
National Institute of Posts and Telecommunication
Rabat, Morocco
ezzarii.mehdi@gmail.com

Hassan Elghazi
STRS Lab
National Institute of Posts and Telecommunication
Rabat, Morocco
elghazi@inpt.ac.ma

Badr Abou El Majd
Department of Mathematics
Faculty of Science
Mohammed V University
Rabat, Morocco
abouelmajd.b@gmail.com

## ABSTRACT
Smart grid is an alternative solution of the conventional power grid which harnesses the power of the information technology to save the energy and meet today's' environment requirements. Due to the inherent vulnerabilities in the information technology, the smart grid is exposed to a wide variety of threats that could be translated into cyber-attacks. In this paper, we develop a deep learning-based intrusion detection system to defend against cyber-attacks in the advanced metering infrastructure network. The proposed machine learning approach is trained and tested extensively on an empirical industrial dataset which is composed of several attack' categories including the scanning, buffer overflow, and denial of service attacks. Then, an experimental comparison in terms of detection accuracy is conducted to evaluate the performance of the proposed approach with Naïve Bayes, Support Vector Machine, and Random Forest. The obtained results suggest that the proposed approaches produce optimal results comparing to the other algorithms. Finally, we propose a network architecture to deploy the proposed anomaly-based intrusion detection system across the Advanced Metering Infrastructure network. In addition, we propose a network security architecture composed of two types of Intrusion detection system types, Host and Network-based, deployed across the Advanced Metering Infrastructure network to inspect the traffic and detect the malicious one at all the levels.

## 1. INTRODUCTION
Smart grid is an alternative of the conventional power system since it harnesses the power of the communication and the information technology (ICT) in generating, delivering, and consuming the energy power. Unlike the traditional power grid, the smart grid enables the two-way flow of information which creates an automated and widely distributed system that offers new functionalities such as operational efficiency, grid resilience, real-time control, and better integration of the renewable technology. However, the inherent weaknesses in the ICT expose the power grid to a wide variety of threats that can be exploited it and translated it into impactful cyber-attacks. Such cyber-attacks can cause an interruption in the power generation causing a disturbance in the system stability and a potential socio-economic impact [10].

Despite the wide variety of industrial cyber-attacks, they usually attempt to compromise these three security parameters: confidentiality, availability, and integrity [11]. The first class of attacks which target the confidentiality includes attacks such as Traffic analysis attack [17], Modbus network scanning [2], and DNP3 network scanning [7] which target the two industrial protocol Modbus and DNP3, respectively. The second category of attacks which target the availability of the system includes attack such as Puppet attack [25], the Time delay switch attack and the Time synchronization attack [27] which are considered as a denial of service attack. The third category of attacks includes the false data injection attack, popping the Home Machine Interface (HMI) [17], Masquerade attack [17], and jamming attack [13].

Defending against these various cyber-attacks in a distributed and heterogeneous system, such as smart grid, is challenging. The smart grid consists of several system and protocols including the supervisory and control and data acquisition (SCADA), the demand response system, the automation substation, and the advanced metering infrastructure (AMI). In this paper, we will focus on defending against cyber-attacks in the AMI network because it constitutes the main connection point between the Home Area Network (HAN) with the Neighborhood Area Network (NAN) and the Wide Area Network (WAN). The AMI Advanced metering infrastructure (AMI) is responsible for collecting, measuring and analyzing energy, water, and gas usage. It allows two-way communication from the user to the utility. It consists of three components: smart meter, AMI headend, and the communication network. Smart meters are digital meters, consisting of microprocessors and a local memory, and they are responsible first for monitoring and collecting power usage of home appliances, and also for transmitting data in real time to the AMI, which is an AMI server consisting of a meter data management system (MDMS).

The communication between the smart meters, the home appliances, and the AMI headend is defined through several communication protocols such as Z-wave and Zigbee [9, 10, 12, 17].

To protect the AMI network and defend against the malicious anomaly, we propose in this paper an intrusion detection system (IDS). There are three distinct categories of IDS: specification-based approach, anomaly-based approach, and signature-based approach. The first approach is based on the logical specification for identifying deviations from the normal behavior profile. The second approach utilizes statistical measures to distinguish between normal and anomalous behavior. The third category consists of discriminating the malicious behavior from the normal one using a database of known attack signatures [3]. The main drawback in adopting the specification and logical-based approaches is that they require a frequent update of the attacks' database and the specifications' list. In this paper, we propose an anomaly-based intrusion detection system (IDS).

A well-known criticism of the anomaly based IDS is its high rate of false positive and low rate of the accuracy. Therefore, several research papers have been published to address these issues and improve the IDS performances in the AMI network. For instance, the authors of [6] proposed an IDS based on both the anomaly-based and signature-based approaches. Their proposed IDS is divided into two main modules: passive and active modules. The passive module is responsible for inspecting the streaming data using the specification based approach while the active module is utilized the anomaly based approach. Their solution is deployed using a single-pass algorithm called FP-Steam. Authors in [20] proposed and a sequence mining based IDS approach. In their approach, the sequence mining algorithms are utilized to collect the network data stream and then a sliding window technique is used for detecting the network intrusions. Based on their obtained results, the proposed approach report high accuracy rate. In [26], authors proposed a distributed IDS for the entire Smart Grid network based on two clonal algorithms named AIRS2Parallel and CLONALEG, which are derived from the support vector machine and the artificial immune system. The proposed IDS architecture is deployed in the grid through three main networks: HAN, NAN, and WAN. The authors of [21] proposed a signature-based IDS approach using the time-table-joined frequent serial episode for extracting the signatures from the time-series stream. In [16], authors proposed a Fuzzy logic approach for detecting automatically the characteristics of clusters, which could be anomalous or a normal cluster, from the network data stream. In [12], the authors proposed an IDS based on the data stream mining approach for the AMI network. In their approach, the authors conducted a comparison performance of seven stream data mining classifiers based on several metrics including the accuracy, the processing time, and the false positive rate.

In our previous paper, we have conducted a performance comparison between several machine learning based IDS approaches including Naïve Bayes, Decision Tree, and Random Forest. To the author's best knowledge, this is the first attempt to propose Deep learning (DL) based approach for the AMI network. In this paper, we propose a feasible security network architecture to deploy the proposed IDS across the AMI network at the key points in order to detect the intrusion at all the levels. The proposed DL-based IDS approach is trained and tested extensively on the NSL KDD data set which includes different categories of cyber attacks where three activation functions are explored: Sigmoid, Relu, and Tanh in order to improve detection accuracy. The optimal output is compared against Support Vector Machine, Naïve Bayes, and Random Forest. The comparison is based on several performance metrics including the probability of detection, the probability of false alarm, and the probability of miss detection.

This paper is divided into three main sections. The first section is divided into three subsections. Subsection one describes how the proposed IDS architecture can detect the intrusions at the key position of the AMI network. The second one explains the DL-based while the third one describes presents the proposed methodology used to preprocess the dataset. The second section presents and discuss the obtained results. In the last section, we draw some conclusions.

## 2. Methodology
## 2.1 Proposed IDS for the AMI network

A typical Advanced Metering Infrastructure (AMI) is composed of three main components: smart meters, an AMI headend, and network communication. A smart meter is a digital meter responsible for collecting in real-time the measurement and transmit these data to AMI headend. The headend consists of two main parts: the AMI server, which is responsible for collecting the meter data, and a meter data management system (MDMS), which manages collected data and shares it with other systems such as demand response systems, historians, billing systems. The communication between the smart meters and the AMI headend passes through several network protocols including the power line communications (PLC), Zigbee, Z-Wave, and the Wireless M-Bus [3].

Since the AMI system is a highly connected network and weaknesses in one component exposes the entire system to a wide variety of risk. Thus, we proposed an end-to-end security strategy with two lines of defense. As shown in Figure. 1, the first line is ensured by the Host Intrusion Detection System (HIDS). The HIDS is deployed at smart meters and the AMI backend server and its main purpose is to protect the firmware, the operating system and the software, and the network interfaces of these devices. There are several cyber attacks can target primarily the smart meter devices and the AMI backend server including virus, worm, Trojan horse, MITM, and DoS attacks. The HIDS defends against those cyber attacks by analyzing specific host-based actions, such as the used applications, the accessed files, and the data resides in the kernel logs.

The second line of defense consists of the Network Intrusion Detection System (NIDS). The AMI network is subject to various network attacks including the Replay attack, Jamming channel attack, and Distributed Denial of Service (DDoS). The NIDS defends against those cyber attacks by sniffing and inspecting the AMI network traffic and it provides a broader examination of the entire network. The HIDS is not able to detect these malicious intrusions until they breach the system while the NIDS can detect the unauthorized intrusion before the attacker reaches the system. Therefore, combining the HIDS and NIDS as two lines of defenses can create a robust defensive strategy to protect the three main AMI components: smart meters, the AMI headend, and the communication network. In this work, we propose a DL-based approach for both the HIDS and NIDS.

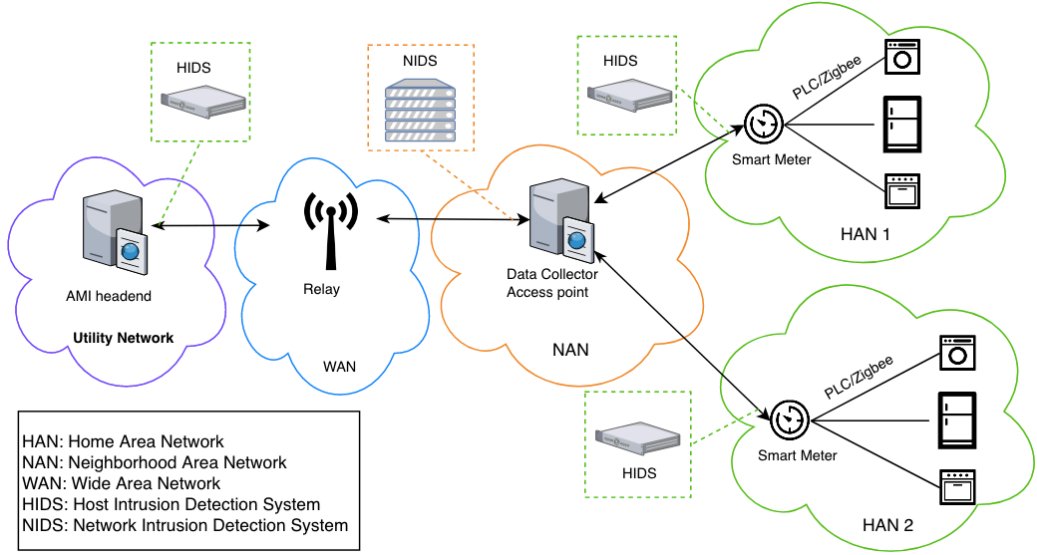

Figure 1. Intrusion detection system architecture for the AMI network

## 2.2 Deep learning based intrusion detection approach

Classifying the events in the AMI network into malicious and normal events can be approached as binary classification problem in which two classes of events are considered: a normal event or malicious event. Assigning each event into one of these two classes requires a supervised machine learning approach which is trained on a labeled dataset. Deep learning (DL) is considered a classification algorithm consisting of an input layer, several hidden layers, usually more than 3 hidden layers, and an output layer. Each layer is composed of several neurons nodes and every node in a layer is directly connected to every other node in the next layer leading to create a deeper neural network [14]. This architecture is used to extract automatically the features from data and make predictive classification about new data.

A typical neuron takes several inputs $x_1, x_2, ..., x_n$ each of which is multiplied by a given weight, $w_1, w_2, ..., w_n$. These inputs are multiplied by their corresponding weights and summed together to pass them through a non-linear activation function. The equation of a given node is expressed as:

$$z = f(b + x * w) f(b + \sum_{i=0}^{n} w_i x_i) \quad (1)$$

Where $b$ is the bias term which allows to shift the results of the activation function to the left or right and to train the model when the input features are 0. In this paper, the inputs $x_1, x_2, ..., x_n$ are the features which represents the each network event in the dataset. Example of these features include the network protocol type, the flag, the service, the IP address of the sender and the receiver, and the Port source and destination. These inputs are fed into an activation function. There are three major types of activation functions: Sigmoid, Relu function, and Tanh function [4].

In Sigmoid function, the output is bounded between 0 and 1. When the input is very large, the output is close to 1, and when the input is very small then the output is close to 0. The values between these two extremes are represented as an S-shape. This function is expressed as [4]:

$$f(z) = \frac{1}{1+e^{-z}} \quad (2)$$

Where $z$ is the input value of a given feature.

The Relu function which stands for the restricted linear unit (ReLu), is a non-linear function expressed as follows [4]:

$$f(z) = \max(0, z) \quad (3)$$

Where $z$ is the input value of a given feature.

The Tanh function is similar to the sigmoid function, however instead of ranging from 0 to 1, the Tanh's output ranges from -1 to 1. The Tanh function is preferred over the sigmoid function since it is zero-centered. Tanh function is expressed as [4]:

$$f(z) = \tanh(0, z) \quad (4)$$

Where $z$ is the input value of a given feature. In this paper, the three types of activation functions are considered to train and the test the proposed model.

Training the DL model passes through two majors steps: forward propagation and backpropagation. In the forward propagation, the training inputs examples are weighted and pass through the activation function to compute an output for each node. This output is compared against the actual output of the training dataset in order to measure the error using the loss function. There is various loss function types including the mean square error, mean absolute error, mean bias error, and Cross Entropy loss function. Since we are dealing with a classification problem, we select the Cross Entropy loss function which is expressed as [19]:

$$CrossEntropyLoss = -(y_i \log(y'_i) + (1 - y_i)\log(1 - y'_i)) \quad (5)$$

Where the $y_i$ is the actual output and $y'_i$ is the computed one. Once the loss function is computed, the backpropagation is performed in order to propagate the error to every node in the network and reducing the error by updating the weights using a gradient descent optimization algorithms. There are several types of optimizers including Adaptive moment estimation (Adam), Nesterov accelerated gradient, Adagrad, Adadelta, RMSprop [23]. According to [23], Adam is the optimal optimizer comparing to other gradient descent optimization algorithms. Thus, Adam is selected as an optimizer with DL in this paper.

Since it is challenging to classify each packet as a normal of an attack with 100% confidence, we use the Softmax function in the last hidden layer to compute the probability of distribution over a set of mutually exclusive labels (0 and 1 in this case). This probability provides the level of confidence of the prediction. The Softmax function is expressed as:

$$y_i = \frac{e^{Z_i}}{\sum_{j=0}^{k} e^{Z_j}} \quad (6)$$

Where $Z_i$ is the activation function of a neuron $i$ and $k$ is the total number of hidden neurons.

## 2.3 NSL_KDD Dataset Preprocessing

The dataset used in this paper is the NSL_KDD dataset which is an improved version of the KDD cup 99 which was gathered at the MIT Lincoln Laboratory [24, 28]. Although the KDD cup 99 and NSL_KDD includes attacks gathered from a conventional network which uses the TCP/IP protocols, it can be used for assessing IDS in an industrial network because most of the Smart Grid network protocol, including DNP3, IEC 61850, IEC 60870, and Modbus uses the TCP/IP protocol. In the NSL_KDD version, several improvements have been made including eliminating the redundant records and removing the bias from the data distribution [5, 18]. The NSL_KDD includes 148 481 records divided into training and testing data. Each record represents a network packet and it consists of 41 features, 34 are numerical and 7 are symbolic, and an output which could be a normal packet or an attack. An attack can belong to one of these four attack's category: Denial of service (DOS), Remote to Local (R2L), Remote to Local (R2L), and Probing [8, 22]. Table 1 gives a description of each attack's category and the number of records of each category in the dataset. In this paper, we put these four attack's categories as one category named: attack. Thus, the dataset has two output either normal or attack packets.

**Table 1. NSL_KDD dataset description**

| Record's type | Description | Examples | Training samples | Testing samples |
|---|---|---|---|---|
| DOS | Denial of service | Teardrop attack and Smurf attack | 45 927 | 7 458 |
| Remote to Local (R2L) | Unauthorized access from a remote machine | Password guessing | 942 | 1 656 |
| User to Root (U2R) | Unauthorized access to local root privileges from a local unprivileged user | Rootkits, buffer overflow attack | 105 | 1 298 |
| Probing | Surveillance and scanning | Scanning attack | 11 656 | 2 421 |
| Normal | Normal packet | TCP SYN, TCP ACK | 67 343 | 9 711 |

Before feeding the DL with the training dataset, the NSL_KDD dataset has to be preprocessed by transforming and standardizing the data in order to help the DL algorithm to converge quickly. Data transformation step consists of converting the nominal and categorical data into numerical data. Table 2 is used as a model to perform a consisting mapping between the nominal symbols and the numerical numbers through the entire dataset.

**Table 2. Converting features nominal values into numerical ones**

| Packet's field | Feature name | Numeric values |
|---|---|---|
| Output | Normal | 0 |
| | Attack | 1 |
| Protocol_type | TCP | 2 |
| | UDP | 3 |
| | ICMP | 4 |
| Flag | OTH | 5 |
| | REJ | 6 |
| | RSTO | 7 |
| | RSTR | 8 |
| | S0 | 9 |
| | S1 | 10 |
| | S2 | 11 |
| | S3 | 12 |
| | SF | 13 |
| | SH | 14 |
| Service | All services | 15 to 80 |

Standardization consists of centering the data around 0 and scaling it with respect to standard variation. The standardization equation is expressed as:

$$x_{standarization} = \frac{x - \mu}{\sigma} \quad (7)$$

Where $x$ is the feature's value, $\mu$ and $\sigma$ are the mean and standard deviation of the dataset, respectively.

## 3. Results and discussion

The proposed DL is implemented using the Tensorflow [1] and Keras framework [15] and is trained and tested using the cross-validation technique with 10 folds. In order to evaluate its performance, two metrics are selected: detection accuracy and a loss function. The accuracy is expressed as the number of correctly classified events over the total number of events. The loss function is given by equation (5). Training the DL for the first time requires some initial parameters, in this paper, we initiate the number of layers to two hidden layers with 300 neuron nodes each and set the number of epochs to 3. Simulation results are given in Figures 2 through 6.

Figure 2 illustrates the accuracy as a function of the number of packets for three different activation functions Relu, Sigmoid, and Tanh functions. As it can be seen, DL with Sigmoid and Softmax function reports the higher accuracy rate followed by Thanh function then the Relu function. The accuracy of the Sigmoid function increases slightly as the number of packets increases to reaches a peak value of 98,34% with 100 768 packets. Similarly, the Tanh function's accuracy follows almost the same trends by increasing slightly as the number of packets increases to reach the highest accuracy value of 94,78% with 125 960 packets. Regarding the Relu's accuracy, it increases and decreases randomly with no tendency. It higher accuracy value is 97,59% reached with 113 364 packets.

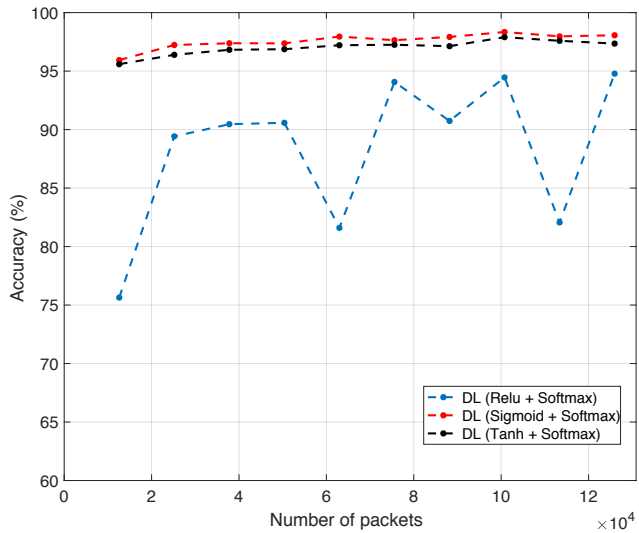

Figure 2. Accuracy as a function of the number of packets for Sigmoid, Tanh, and Relu activation functions

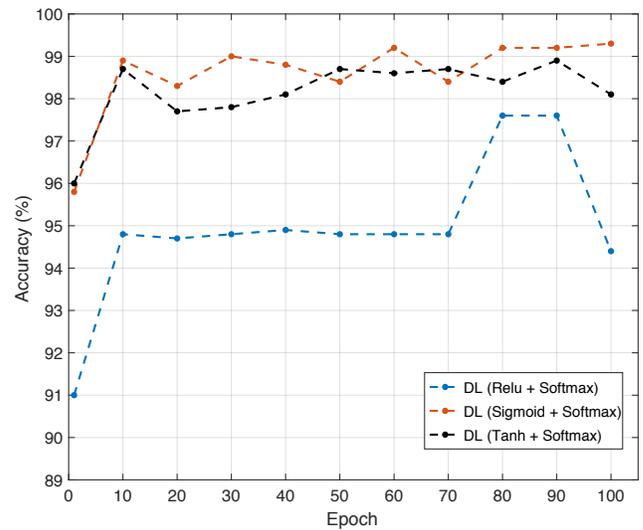

Figure 4. Accuracy as a function of the number of epochs for Sigmoid, Tanh, and Relu activation functions

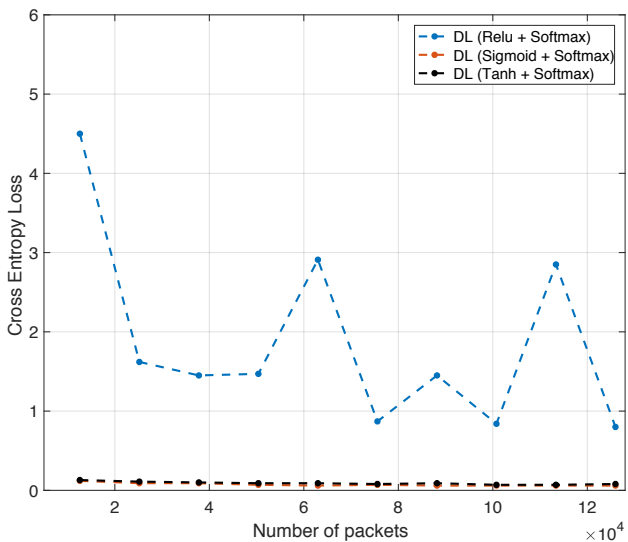

Figure 3. Cross Entropy Loss as a function of the number of packets for Sigmoid, Tanh, and Relu activation functions

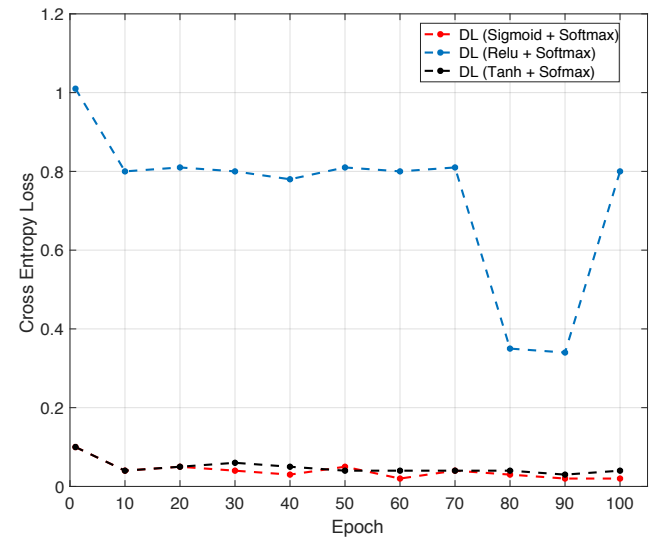

Figure 5. Cross Entropy Loss as a function of the number of epochs for Sigmoid, Tanh, and Relu activation functions

Figure 3 depicts the cross entropy loss function as a function of the number of packets. As one can see, the Sigmoid with the Softmax function reports the lower loss in overall followed by the Tanh with function and then Relu function. The loss of Sigmoid and Tanh remains almost the same for a different number of packets. With Sigmoid, it reaches the lowest value of 0,06 with 62980 packets. Tanh's reaches the lowest accuracy, which is 0,07 with 100 768 packets. The loss of the Relu function does not show any tendency, it decreases and then increases randomly with the lowest value of 0,8. The results of Figures 2 and 3 suggest that the optimal number of packets for training and testing the DL is 125 960. In addition, the Sigmoid function is the optimal activation function to improve the DL performance in terms of accuracy and error.

Figure 4 shows the accuracy of the three activation functions with Softmax as a function of the number of epochs. An epoch means an entire pass through the training dataset. It is essential to define the optimal number of epochs required for the gradient descent in order to converge to an optimum point. As it can be seen, the accuracy of the Sigmoid function with the Softmax function increases between 2 and 10 epochs then it changes slightly between 98% and 99.5% before it stabilizes at 80 epochs and reaches the highest accuracy values of 99.5% at 100 epochs. Similarly, The Tanh with the Softmax function follows the same tendency with lower accuracy between 5 and 40 epochs and it outperforms the Sigmoid function when the number of epochs is equal to 50 and 70 epochs. The Relu with the Softmax function exhibits low accuracy. It increases exponentially between 2 and 10 epochs and it stabilizes at 94% before increasing to 97% with 71 epochs and then drops to 94% at 100 epochs.

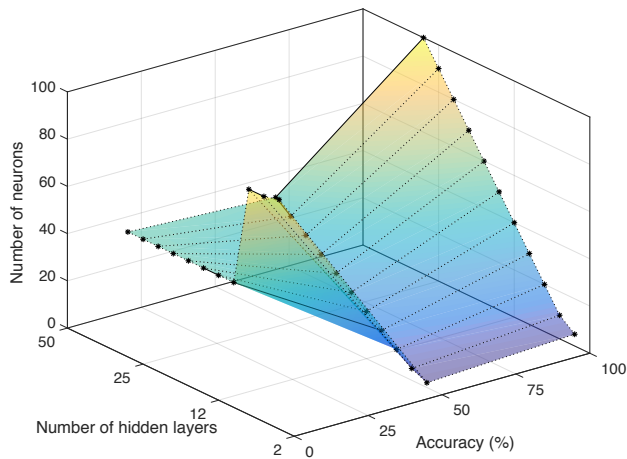

Figure 6. The accuracy of the DL with Sigmoid function as a function of the number of layers and the number of hidden nodes.

Figure 5 illustrates the cross entropy loss as a function of the number of epochs for Deep learning with three different activation functions: Sigmoid, Relu, and Tanh. As it is expected, the DL with Sigmoid function exhibits the lowest error rate followed by Tanh and then Relu function. As it can be seen, the Sigmoid function changes slightly as the number of epochs increases to reach the lowest error value of 0,02 at 100 epochs. Similarly, the Tanh function changes slightly as the number of epochs changes. It outperforms the Sigmoid function with 50 epochs and it reports the same error value at 70 epochs. Its lowest error value is reached with 90 epochs while the highest one is reached with 30 epochs. The Relu function exhibits poor performance in terms of error comparing to Sigmoid and Tanh. It decreases to 0,8 with 10 epochs and it stabilizes between 10 and 70 epochs and then it drops to 0,34 with 90 epochs before returning to 0,8 with 100 epochs. The results given in Figure 4 and 5 suggest that DL with the Sigmoid activation function with 100 epochs exhibit the optimal performance in terms of accuracy and error. Thus, we conducted a parametric study in order to define the optimal number of nodes and hidden layers to improve further the performance of the DL with the Sigmoid function in terms of accuracy. The obtained results are given in Figure 6. As one can see, the higher the accuracy value, which 97.1% is produced by 2 hidden layers with 5 neuron nodes each. The accuracy decreases to 95.5% with 5 and 10 hidden layers then it drops and stagnates at 52.9% with more than 10 layers.

Table 3 gives the comparison's results of the proposed DL-based approach with three machine learning based IDS approaches namely: Random Forest, Support Vector Machine, and Naïve Bayes. This comparison was conducted on 100 768 packets where the cross validation techniques were used to evaluate the performance of each approach. For DL, we selected the optimal results obtained from the previous experiment by selecting the Sigmoid as an activation function with 2 hidden layers and 5 neurons nodes each. Regarding the other approaches, we conducted a parametric study for each approach in order to select the optimal parameters' values which produce the accurate results. Therefore, we set the number of trees in the Random Forest to 100, and select the Radial Basis Function (RBF) kernel for the SVM. As it can be seen from Table 3, the proposed approach reports the optimal result with a detection accuracy of 95.5%. Random Forest exhibits good performances with an accuracy of 99.3% then Naïve Bayes which reports 88.9% and finally SVM which shows poor performance with an accuracy of 61.7%.

Table 3. Comparison between the proposed approach with Random Forest-based IDS and SVM-based IDS

| Machine learning-based IDS approaches | Accuracy |
|---|---|
| Proposed DL-based IDS approach | 99.5% |
| Random Forest-based IDS | 99.3% |
| SVM-based IDS | 61.7% |
| Naïve Bayes-based IDS | 88.9% |

## 4. CONCLUSION

In this paper, we developed a Deep learning based Intrusion detection system approach for the Advanced Metering Infrastructure Network. The proposed classifier is trained and tested extensively on the NSL_KDD dataset which includes more than 148 480 events, which are classified into normal and attack event, and 41 features. An extensive experimental study is conducted to define the suitable number of hidden layers, number of nodes and the activation function required to classify accurately each event. In addition, the proposed approach is compared against three machine learning based IDS approach: Random Forest, Naïve Bayes, and Support Vector Machine. The obtained results suggest that the proposed DL-based IDS approach with the Sigmoid activation function, 2 hidden layers, and 5 neurons nodes each is able to classify the packets into normal and malicious with an accuracy of 99.5%. The comparison results show that the proposed approach outperforms Random Forest, SVM, and Naïve Bayes based IDS approach. In addition, we proposed a network security architecture to deploy the proposed DL-based IDS at the host and the network key location in order to detect the intrusions at all levels of the AMI network. As an extension of this work, we will set an AMI testbed, simulate some industrial cyber-attacks, and then deploy the proposed IDS approach in order to evaluate its performance in a real AMI network environment.

## 5. REFERENCES

[1] Abadi, M., Barham, P., Chen, J., Chen, Z., Davis, A., Dean, J., Devin, M., Ghemawat, S., Irving, G. and Isard, M. 2016. Tensorflow: a system for large-scale machine learning. *OSDI* (2016), 265–283.

[2] Al-Dalky, R., Abduljaleel, O., Salah, K., Otrok, H. and Al-Qutayri, M. 2014. A Modbus traffic generator for evaluating the security of SCADA systems. *2014 9th International Symposium on Communication Systems, Networks Digital Sign (CSNDSP)* (Jul. 2014), 809–814.

[3] Berthier, R., Sanders, W.H. and Khurana, H. 2010. Intrusion Detection for Advanced Metering Infrastructures: Requirements and Architectural Directions. (Oct. 2010), 350–355.

[4] Buduma, N. and Locascio, N. 2017. *Fundamentals of Deep Learning: Designing Next-Generation Machine Intelligence Algorithms*. O'Reilly Media, Inc.

[5] Chae, H., Jo, B., Choi, S.-H. and Park, T. 2013. Feature selection for intrusion detection using NSL-KDD. *Recent advances in computer science*. (2013), 184–187.

[6] Chu, N.C., Williams, A., Alhajj, R. and Barker, K. 2004. Data stream mining architecture for network intrusion detection. *Information Reuse and Integration, 2004. IRI 2004. Proceedings of the 2004 IEEE International Conference on* (2004), 363–368.

[7] Darwish, I., Igbe, O., Celebi, O., Saadawi, T. and Soryal, J. 2015. Smart Grid DNP3 Vulnerability Analysis and Experimentation. (Nov. 2015), 141–147.

[8] Dhanabal, L. and Shantharajah, S.P. 2015. A study on NSL-KDD dataset for intrusion detection system based on classification algorithms. *International Journal of Advanced Research in Computer and Communication Engineering*. 4, 6 (2015), 446–452.

[9] El Mrabet, Z., Arjoune, Y., El Ghazi, H., Abou Al Majd, B., Kaabouch, N., El Mrabet, Z., Arjoune, Y., El Ghazi, H., Abou Al Majd, B. and Kaabouch, N. 2018. Primary User Emulation Attacks: A Detection Technique Based on Kalman Filter. *Journal of Sensor and Actuator Networks*. 7, 3 (Jul. 2018), 26. DOI:https://doi.org/10.3390/jsan7030026.

[10] El Mrabet, Z., Kaabouch, N., El Ghazi, H. and El Ghazi, H. 2018. Cyber-security in smart grid: Survey and challenges. *Computers & Electrical Engineering*. 67, (2018), 469–482.

[11] Elmrabet, Z., Elghazi, H., Sadiki, T. and Elghazi, H. 2016. A New Secure Network Architecture to Increase Security Among Virtual Machines in Cloud Computing. *Advances in Ubiquitous Networking*. E. Sabir, H. Medromi, and M. Sadik, eds. Springer Singapore. 105–116.

[12] Faisal, M.A., Aung, Z., Williams, J.R. and Sanchez, A. 2015. Data-stream-based intrusion detection system for advanced metering infrastructure in smart grid: A feasibility study. *IEEE Systems Journal*. 9, 1 (2015), 31–44.

[13] Gai, K., Qiu, M., Ming, Z., Zhao, H. and Qiu, L. 2017. Spoofing-Jamming Attack Strategy Using Optimal Power Distributions in Wireless Smart Grid Networks. *IEEE Transactions on Smart Grid*. (2017), 1–1. DOI:https://doi.org/10.1109/TSG.2017.2664043.

[14] Goodfellow, I., Bengio, Y., Courville, A. and Bengio, Y. 2016. *Deep learning*. MIT press Cambridge.

[15] Gulli, A. and Pal, S. 2017. *Deep Learning with Keras*. Packt Publishing Ltd.

[16] Khan, M.U. 2009. Anomaly detection in data streams using fuzzy logic. *Information and Communication Technologies, 2009. ICICT'09. International Conference on* (2009), 167–174.

[17] Knapp, E.D. and Samani, R. 2013. *Applied cyber security and the smart grid: implementing security controls into the modern power infrastructure*. Elsevier, Syngress.

[18] Lakhina, S., Joseph, S. and Verma, B. 2010. Feature reduction using principal component analysis for effective anomaly–based intrusion detection on NSL-KDD. (2010).

[19] Lin, Y., Lin, H., Wu, J. and Xu, K. 2011. Learning to Rank with Cross Entropy. *Proceedings of the 20th ACM International Conference on Information and Knowledge Management* (New York, NY, USA, 2011), 2057–2060.

[20] Liu, Q.-H., Zhaoi, F. and Zhao, Y.-B. 2005. A real-time architecture for NIDS based on sequence analysis. *Machine Learning and Cybernetics, 2005. Proceedings of 2005 International Conference on* (2005), 1893–1896.

[21] Qun, Z. and Wen-jie, H. 2010. Research on data mining technologies appling intrusion detection. *Emergency Management and Management Sciences (ICEMMS), 2010 IEEE International Conference on* (2010), 230–233.

[22] Revathi, S. and Malathi, A. 2013. A detailed analysis on NSL-KDD dataset using various machine learning techniques for intrusion detection. *International Journal of Engineering Research and Technology. ESRSA Publications*. (2013).

[23] Ruder, S. 2016. An overview of gradient descent optimization algorithms. *arXiv:1609.04747 [cs]*. (Sep. 2016).

[24] Tavallaee, M., Bagheri, E., Lu, W. and Ghorbani, A.A. 2009. A Detailed Analysis of the KDD CUP 99 Data Set. *Proceedings of the Second IEEE International Conference on Computational Intelligence for Security and Defense Applications* (Piscataway, NJ, USA, 2009), 53–58.

[25] Yi, P., Zhu, T., Zhang, Q., Wu, Y. and Li, J. 2014. A denial of service attack in advanced metering infrastructure network. *2014 IEEE International Conference on Communications (ICC)* (2014), 1029–1034.

[26] Zhang, Y., Wang, L., Sun, W., Green II, R.C. and Alam, M. 2011. Distributed intrusion detection system in a multi-layer network architecture of smart grids. *IEEE Transactions on Smart Grid*. 2, 4 (2011), 796–808.

[27] Zhang, Z., Gong, S., Dimitrovski, A.D. and Li, H. 2013. Time Synchronization Attack in Smart Grid: Impact and Analysis. *IEEE Transactions on Smart Grid*. 4, 1 (Mar. 2013), 87–98. DOI:https://doi.org/10.1109/TSG.2012.2227342.

[28] The KDD99 dataset available at: https://kdd.ics.uci.edu/databases/kddcup99/task.html.